\documentclass[12pt]{article}
\usepackage{graphicx}
\usepackage{amsmath,amssymb}
\usepackage{float}
\usepackage{cite}

\newcommand{\p}{\bot}

\newcommand{\de}{\delta}

\newcommand{\e}{\varepsilon}
\newcommand{\ls}{\left(}
\newcommand{\rs}{\right)}

\newcommand{\g}{\gamma}
\newcommand{\ff}{\varphi}
\newcommand{\la}{\lambda}
\newcommand{\m}{\mu}

\newcommand{\disn}[2]{$$\displaylines{\refstepcounter{equation}\label{#1}\hskip 1em minus 1em #2\hfilneg}$$}
\newcommand{\nom}{\hfil\hskip 1em minus 1em (\theequation)}
\newcommand{\ns}{\hfill\cr\hfill}

\begin{document}

\title{Pauli-Villars Regularization in nonperturbative Hamiltonian approach on the Light Front}

\author{
M.Yu.~Malyshev\thanks{E-mail: mimalysh@yandex.ru},
S.A.~Paston,
E.V.~Prokhvatilov,
R.A.~Zubov,
V.A.~Franke\\
{\it Saint Petersburg State University, St.-Petersburg, Russia}
}
\date{\vskip 15mm}
\maketitle

\begin{abstract}
The advantage of Pauli-Villars regularization in quantum field theory quantized on the light front is explained. Simple examples of scalar $\la\varphi^4$ field theory and Yukawa-type model are used. We give also an example of nonperturbative calculation in the theory with Pauli-Villars fields, using for that a model of anharmonic oscillator modified by inclusion of ghost variables playing the role similar to Pauli-Villars fields.
\end{abstract}



\maketitle

\section{Introduction}
In this paper we consider Pauli-Villars (PV) regularization \cite{PV} not only as a way of Lorentz invariant ultraviolet (UV) regularization in field theory but also as a way of restoring of the equivalence between perturbation theory, related to quantization on the light front (LF), and usual covariant perturbation theory in Lorentz coordinates \cite{NPPF}. Owing to such advantage this regularization can be applied to the construction of renormalized LF Hamiltonian which can be used for nonperturbative calculations \cite{paul,
Bakker.Bassetto.Brodsky.Broniowski.Dalley.Frederico.Glazek.Hiller.Ji.Karmanov.et.al.arXiv2013}.

Let us remind the formulation of field theory in the light front (LF) coordinates:
\disn{I1}{
x^{\pm} = \frac{x^0 \pm x^1}{\sqrt{2}},\,\, x^{\p},
\nom}
where $x^0,\, x^1,\, x^{\p}$ are Lorentz
coordinates, $x^+$ plays the role of time and $x^-$ plays the role of space coordinate. These coordinates were introduced by P.A.M. Dirac \cite{dir}. The quantization on the LF means the quantization on the plane $x^+=0$.
The generator $P_-$ of translations in $x^-$ is kinematical \cite{dir} (i.e. it is independent of the
interaction and quadratic in fields, as a momentum in a free theory). It is nonnegative ($P_-\geqslant0$)
for quantum states with nonnegative mass squared. So the state with the minimal eigenvalue $p_-=0$ of the
momentum operator $P_-$ can describe (in the case of the absence of the massless particles) the vacuum
state, and it is
also the state minimizing the $P_+$ in Lorentz invariant theory.
This can simplify the description of vacuum state in quantum field theory.
Furthermore it is possible
to introduce the Fock space on this vacuum and  formulate in this space the eigenvalue problem for the
operator $P_+$
(which is the LF Hamiltonian) and find the  spectrum of mass $m$ in subspaces with fixed values of the
momenta $p_-, p_{\bot}$ \cite{NPPF, paul}\footnote{Review \cite{NPPF} includes the content of papers \cite{Ilgen, tmf97, tmf99, tmf02}.}:
\disn{1aa}{
P_+|p_-,p_{\bot}\rangle=\frac{m^2+p_{\bot}^2}{2p_-}|p_-,p_{\bot}\rangle.
\nom}

The theory on the LF has the singularity at $p_- = 0$,
and the simplest regularization is the cutoff $p_-\geqslant\e>0$.
Other convenient translationally invariant regularization, that can treat also
 zero ($p_-=0$) modes of fields, is the cutoff $|x^-| \leqslant L$ plus
periodic boundary conditions for fields. This regularization
discretizes the momentum $p_-$ ($p_-=\frac{\pi n}{L} ,\,
n=0,1,2,...$) and clearly separates zero and nonzero modes. It is
the so-called "Discretized Light Cone Quantization" (DLCQ). Such
regularization was successfully used to solve the problem
(\ref{1aa}) for (1+1) field theories
\cite{Annenkova, Brodsky.Pauli, Eller.Pauli.Brodsky, Hornbostel.Brodsky.Pauli}.
Nevertheless the problem  of constructing the renormalized LF
Hamiltonians using, in particular, the above-mentioned
regularizations turned out to be very difficult. We refer to
nonperturbative "similarity renormalization group" (SRG) approach
which
allows to construct approximately effective LF Hamiltonians acting
in the space of small number of effective (constituent) particles
\cite{Wilson, Glazek, Glazek.Wilson.Phys.Rev.D1993,
Glazek.Wilson.Phys.Rev.D1994, Glazek.ActaPhys.Polon.B2008, Glazek.Phys.Rev.D2001, Glazek.Wieckowski.Phys.Rev.D2002}.

All used regularizations of the singularity at $p_- = 0$ are not Lorentz invariant.
This can lead to nonequivalence of the results
obtained with
the LF and the conventional  formulation in Lorentz coordinates.
It was shown in papers \cite{NPPF, tmf97, burlang, Burkardt.Langnau} that some diagrams
of the perturbation theory, generated by the LF Hamiltonian, and
corresponding diagrams of the conventional  perturbation theory in
Lorentz coordinates can differ.
In  papers \cite{NPPF, tmf97} it was found how to restore the equivalence of the LF and
conventional perturbation theories in all orders in the coupling constant by addition of new
(in particular, nonlocal) terms to the canonical LF Hamiltonian. These terms must remove the
above-mentioned
differences of diagrams.

The method of the restoration of the equivalence between the LF and
conventional perturbation theories, found in \cite{NPPF, tmf97}, was applied to constructing of correct
renormalized LF Hamiltonian for (3+1)-dimensional Quantum Chromodynamics \cite{NPPF, tmf99}.
In the papers \cite{NPPF, tmf02} this method was applied to massive Schwinger model
((1+1)-dimensional
Quantum Electrodynamics) and correct LF Hamiltonian was constructed. This Hamiltonian was used for
numerical calculations
of the mass spectrum \cite{Yad.Fiz.2005}, and the obtained results well agree with lattice
calculations in Lorentz coordinates \cite{Hamer2000}
for all values of the coupling (including very large ones).

The number of the above-mentioned new terms, which must be added to canonical LF Hamiltonian, and
counterterms,
necessary for the ultraviolet (UV) renormalization,
depends essentially on the  regularization scheme. For the case of QCD(3+1)
 \cite{NPPF, tmf99}  in the light-cone
gauge one gets the finite number of these terms only in the regularization of the Pauli-Villars (PV)
 type \cite{PV}.

 This regularization violates gauge invariance. However it was shown in \cite{NPPF, tmf99}
that  gauge invariance can be restored in  renormalized  LF theory with proper choice of coefficients
before these new terms and counterterms.
 On the other side, the PV regularization involves the introduction of auxiliary ghost fields (with
the large  mass playing the
role of the regularization parameter). These ghost fields generate the states with
the indefinite metric, and one has to deal with such states in
the nonperturbative (e.g. variational) calculations using the LF Hamiltonian.
Attempts to do these calculations were made in papers
\cite{Brodsky1, Brodsky2, Brodsky3, Brodsky4} for nongauge theories.
It is important to generalize this for gauge theories like QCD (e.g. for the formulation
 \cite{NPPF, tmf99}, where the PV regularization introduces ghost gauge fields).

The question of using the PV regularization in the
LF Hamiltonian approach isn't studied sufficiently.
So we address this question in the present paper.

In Section~\ref{scalar} we remind the idea of the PV regularization. Then in Section~\ref{restoring the equivalence} we explain the role of this regularization
in restoring the equivalence between perturbation theories related to usual and LF quantization. We show
how the introduction of additional PV fields can remove the difference between results of calculations of Feynman diagrams in these formulations of perturbation theory. Also, as an example of the theory with the PV-like ghosts, in Section~\ref{Anharmonic.oscillator} we consider the modification of anharmonic oscillator model with ghost variables. We calculate the spectrum in this model and show how  different Hamiltonian eigenvalues approach to the eigenvalues of the usual anharmonic oscillator when the parameter playing the role of large PV mass goes to infinity.

\section{Example of Pauli-Villars regularization in scalar field theory}
\label{scalar}
Before explaining the role of PV regularization for perturbative renormalization of the theory quantized on the LF let us remind in this section the idea of PV regularization as a conventional UV regularization. In particular, the PV regularization can be applied for UV regularization of the LF Hamiltonian. We consider, as an example, the scalar field theory with the interaction $\la \varphi^4$  ($\la$ is the coupling constant). Lagrangian density of this theory is the following:
\disn{S1a}{
{\cal{L}} =\frac{1}{2}\partial_{\m}\ff\partial^{\m}\ff
-\frac{m^2}{2}\ff^2 - \la \ff^4.
\nom}
This theory requires UV regularization (e.g. in (2+1) and (3+1) dimensions).
Let us show that it is sufficient to introduce one extra field with large mass playing the role of the regularization parameter. Lagrangian density can be rewritten as follows:
\disn{S1b}{
{\cal{L}} =\sum_{l=0}^1
(-1)^l\ls\frac{1}{2}\partial_{\m}\ff_l\partial^{\m}\ff_l
-\frac{m_l^2}{2}\ff_l^2\rs - \la :\ff^4:,\quad
\ff=\sum_{l=0}^1\ff_l.
\nom}
Here we introduce the symbol "::" which means that in perturbation theory we drop Feynman diagrams with loops containing only one vertex \cite{Weinberg}. The $\ff_0$ is usual field with mass $m_0$ and the $\ff_1$ is the extra field with mass $m_1$.

It can be shown that in (3+1)-dimensional theory the introduction of one PV field is sufficient for the ultraviolet regularization (of perturbation theory in $\la$) \footnote{We discard Feynman diagrams containing loops with only one vertex. Otherwise we need one more PV field.}.
One can show that momentum space Feynman diagrams in the original theory (\ref{S1a}) diverge no more than quadratically \cite{Pierre.Ramond} (beside of vacuum diagrams).
If we consider Feynman diagrams in the theory (\ref{S1b})
we see that propagators of fields $\ff_0$ and
$\ff_1$ sum up in corresponding diagrams so that
we obtain the following expression which plays
the role of regularized propagator:
\disn{S2}{
\Delta(k)=\frac{i}{k^2-m_0^2+i0}-\frac{i}{k^2-m_1^2+i0}=
\frac{i(m_0^2-m_1^2)}{(k^2-m_0^2+i0)(k^2-m_1^2+i0)}, \ns k^2=k_0^2-\overrightarrow{k}^2 \quad \text{in Lorentz coordinates}.
\nom}
The power of momenta in the denominator of this propagator is four while in the original propagator it is only two. The power four is sufficient for the convergence of integrals corresponding to Feynman diagrams in this theory at finite parameter $m_1$.
Let us remark that extra field $\ff_1$ (ghost field) enters into Lagrangian density (\ref{S1b}) with the opposite sign than the field $\ff_0$.
This leads to the unusual commutation relations for creation and annihilation operators corresponding to this field and creates the states with indefinite metric.

Thus we see that introduction of ghost fields in a way proposed by W.~Pauli and F.~Villars can give UV convergence of Feynman integrals necessary for their UV regularization. In the next Section we consider the advantage of PV regularization for a field theory quantized on the LF.

\section{The role of Pauli-Villars regularization in restoring the equivalence between light front and conventional formulation of field theory}
\label{restoring the equivalence}
As mentioned in the introduction the PV fields can remove the difference between results of calculations of Feynman diagrams of usual covariant
perturbation theory and analogous diagrams  generated by quantization on the LF (LF perturbation theory). In LF perturbation theory we use the cutoff $|p_-|\geqslant\e$ while this cutoff is absent in covariant perturbation theory in Lorentz coordinates. The above mentioned difference is related to the contribution of the domain $|p_-|\leqslant\e$. One can prove that for the most of Feynman diagrams this difference disappears in the $\e\to0$ limit if sufficient number of PV fields are included into the theory, and such limit must be considered at fixed masses of PV field. \footnote{This difference still remains for diagrams with two external lines joined to one vertex.}
Let us explain this using Yukawa-type model as an example. Lagrangian density is the following:
\disn{Y0a}{
{\cal{L}} =\frac{1}{2}\partial_{\m}\ff^a\partial^{\m}\ff^a
-\frac{m^2}{2}\ff^a\ff^a - \la (\ff^a\ff^a)^2 +\overline{\psi}(i\g^{\m}\partial_{\m}-M)\psi-
g\ff^a\overline{\psi}\tau_a\psi,
\nom}
where $\psi$ is fermion field with mass $M$, matrices $\g^{\m}$ are Dirac matrices, $\tau_a$ are Pauli matrices ($a=1,2,3$), $g$ and $\la$ are the coupling constants.

\begin{figure}
  \centering
  \includegraphics[height=.3\textheight]{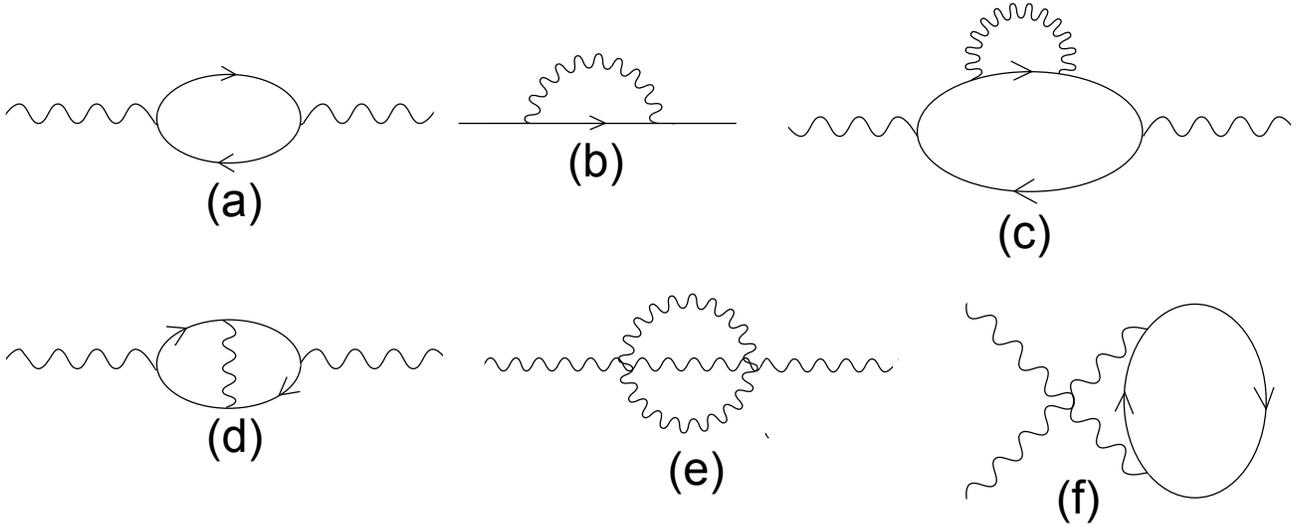}
  \caption{The divergent diagrams of (2+1)-dimensional Yukawa-type model. Wavy lines denotes boson propagators, straight lines denotes fermion propagators.}
\label{diagrams}
\end{figure}

Let us consider only (2+1)-dimensional theory. This theory is superrenormalizable, i.e. it is sufficient to consider finite number of UV divergent diagrams for UV renormalization. These diagrams in Lorentz covariant perturbation theory are shown in Fig.~\ref{diagrams}.
The diagram (a) diverges linearly and the other
diagrams diverge logarithmically in momentum space.

For the UV regularization of these diagrams it is sufficient to introduce one boson and one fermion PV field analogously to the previous example of scalar field. These extra fields lead to effective regularization of the boson and fermion propagators and to UV regularization of divergent diagrams. Lagrangian density including these PV fields has the following form:
\disn{Y0}{
{\cal{L}} =\sum_{l=0}^1(-1)^l\ls\frac{1}{2}\partial_{\m}\ff_l^a\partial^{\m}\ff_l^a
-\frac{m_l^2}{2}\ff_l^a\ff_l^a\rs - \la :(\ff^a\ff^a)^2: +\sum_{l=0}^1(-1)^l\overline{\psi}_l(i\g^{\m}\partial_{\m}-M_l)\psi_l-
g:\ff^a\overline{\psi}\tau_a\psi:,\ns
\ff^a=\sum_{l=0}^1\ff_l^a,\quad\psi=\sum_{l=0}^1\psi_l,
\nom}
where $\psi_0$ is usual fermion field with mass $M_0$, $\psi_1$ is the PV fermion field with the mass $M_1$, the $\ff_0^a$ is usual field with the mass $m_0$, the field $\ff_1^a$ is the PV boson field with the mass $m_1$. As before, the symbol "::" means that in perturbation theory, corresponding this Lagrangian, Feynman diagrams with loops containing only one vertex are absent.

As in the example of scalar field theory we obtain the regularized fermion and boson propagators summing the contribution of similar diagrams, one with usual and the other with PV propagator.
Regularized boson propagator has the same expression (\ref{S2}), as for scalar field theory case (multiplied by $\de^{ab}$). Regularized fermion propagator
has the following expression:
\disn{Yb}{
\Delta_{\psi}(k)=i \,\frac{\g^{\m}k_{\m}+M_0}{k^2-M_0^2+i0}-i \, \frac{\g^{\m}k_{\m}+M_1}{k^2-M_1^2+i0}=i\frac{(M_0-M_1)k^2-
(M_1^2-M_0^2)\g^{\m}k_{\m}-
M_0M_1^2+M_1M_0^2}{(k^2-
M_0^2+i0)(k^2-M_1^2+i0)}.
\nom}

Furthermore we have to remove the above mentioned difference between diagrams in usual and LF perturbation theories. Using the methods of the papers \cite{NPPF, tmf97} we show that by addition of only one boson and one fermion PV field we can remove this difference for all diagrams beside of diagrams with two external lines joined to one vertex because these diagrams are absent in the LF perturbation theory (due to the regularization $|p_-|\geqslant\e$). To get the equivalence between LF and usual perturbation theory in Lorentz coordinates we must introduce into the interaction of usual perturbation theory the counterterm which  cancels these diagrams in perturbation theory and particularly renormalize the theory (\ref{Y0}).

Let us consider the example which illustrates how the above mentioned difference can disappear in
fermion self-energy Feynman diagram (Fig.~1(b)) regularized with the help of PV fields. Integral corresponding to this diagram can be written in the following form:
\disn{Yc}{
\int d^3k\frac{\ls(M_0-M_1)k^2-
(M_1^2-M_0^2)\g^{\m}k_{\m}-
M_0M_1^2+M_1M_0^2\rs\ls m_0^2-m_1^2\rs}{(k^2-
M_0^2+i0)(k^2-M_1^2+i0)((p-k)^2-
m_0^2+i0)((p-k)^2-m_1^2+i0)},
\nom}
where $p$ and $k$ are external and loop momenta respectively. In the calculation of diagrams in usual covariant perturbation theory the integration is carried out
over all momenta $k_{\m}$ while in the
LF calculation the integration is only over the domain
$\{|k_-|\geqslant\e\}\cap \{|k_--p_-|\geqslant\e\}$ due to the regularization
of fields mentioned in the Introduction. So the difference between the LF and the conventional covariant
calculations of the diagram is given by the integral over the domain
$\{|k_-|<\e\}\cup\{|k_--p_-|<\e\}$. This domain consists of two parts and the contribution of
each of them should be considered separately. However the  contribution of the second part becomes
similar to contribution of the first part after the change $k \to \tilde{k} = p-k$. So we consider
the integration only over the first part of the integration domain:
\disn{Y1}{
\int_{-\infty}^{\infty} dk_{\p}\int_{-\infty}^{\infty}dk_+\int_{-\e}^{\e} dk_- \frac{\ls(M_0-M_1)k^2-
(M_1^2-M_0^2)\g^{\m}k_{\m}-
M_0M_1^2+M_1M_0^2\rs(m_0^2-m_1^2)}{(k^2-
M_0^2+i0)(k^2-M_1^2+i0)((p-k)^2-
m_0^2+i0)((p-k)^2-m_1^2+i0)},\ns \quad k^2=2k_+k_--k_{\p}^2, \quad (p-k)^2=2(p_+-k_+)(p_--k_-)-(p_{\p}-k_{\p})^2.
\nom}
After the change $k_- \to\e k_-$, $k_+ \to k_+/\e$ the part of this integral with $\g^+$ takes the form:
\disn{Y2}{
\int_{-\infty}^{\infty} dk_{\p}\int_{-\infty}^{\infty}dk_+\int_{-1}^{1} dk_-\frac{(M_1^2-M_0^2)(m_0^2-m_1^2)\g^+k_+/\e}{(k^2-
M_0^2+i0)(k^2-M_1^2+i0)}\times\ns\times\frac{1}{\ls 2(p_+-k_+/\e)(p_--\e k_-)-(p_{\p}-k_{\p})^2-m_0^2+i0\rs}
\times\ns\times\frac{1}{\ls 2(p_+-k_+/\e)(p_--\e k_-)-(p_{\p}-k_{\p})^2-m_1^2+i0\rs}=\ns=
\e(M_1^2-M_0^2)(m_0^2-m_1^2)\int_{-\infty}^{\infty} dk_{\p}\int_{-\infty}^{\infty}dk_+\int_{-1}^{1} dk_-\frac{\g^+k_+}{(k^2-
M_0^2+i0)(k^2-M_1^2+i0)}\times\ns\times\frac{1}{\ls 2(\e p_+-k_+)(p_--\e k_-)-\e(p_{\p}-k_{\p})^2-\e m_0^2+i0\rs}
\times\ns\times\frac{1}{\ls 2(\e p_+-k_+)(p_--\e k_-)-\e(p_{\p}-k_{\p})^2-\e m_1^2+i0\rs}.
\nom}
The integration domain is now independent on $\e$ while the expression for the integrand can be
analyzed in the $\e \to 0$ limit at fixed values of parameters $m_1$, $M_1$. We see that in this limit the integral (\ref{Y2}) is equal
to zero. It is important that PV regularization must be removed after LF regularization ($m_1,M_1\to\infty$ after $\e \to 0$). Analogously the other parts of integral (\ref{Y1}) and such integrals for other diagrams are equal
to zero in the $\e \to 0$ limit. To see the presence of the difference in the theory without PV fields it is sufficient to consider the original theory (\ref{Y0a}) and cutoff in UV the momenta in the diagram (b).

Thus introducing PV fields we can get the coincidence of LF perturbation theory and covariant perturbation theory in Lorentz coordinates for the most of Feynman diagrams. Adding the counterterm which compensates the remaining differences in Feynman diagrams \footnote{We mean diagrams with two external lines joined to one vertex.} we can get complete coincidence of these perturbation theories. Further we can construct the renormalized LF Hamiltonian for this theory calculating the divergent parts of diagrams shown in Fig.1 and introducing the corresponding counterterms into the original LF Hamiltonian. Such renormalized LF Hamiltonian can be used for nonperturbative calculations. However these calculations can meet difficulties because of the presence of the ghost PV fields and states with indefinite metric. It is interesting to consider this problem in more details. In the next section we give the example of such calculation using the anharmonic oscillator modified by inclusion of ghost variables analogous to PV fields.

\section{Anharmonic oscillator}
\label{Anharmonic.oscillator}
In this Subsection we consider the problem mentioned at the end of previous section, i.e. calculation the spectrum of anharmonic oscillator modified by ghost variables analogous to PV fields. The Hamiltonian of this model is the following:
\disn{1}{
H = \omega_a a^+ a - \omega_b b^+ b + \lambda (a + b + a^+ + b^+)^4,
\nom}
with the parameters $\omega_a$ and $\omega_b$, $\omega_b \gg \omega_a$. The operators $a, a^+, b, b^+$ satisfy the following commutation relations:
\disn{1a}{
[a,a^+] = - [b,b^+] = 1, \,\, [a,a] = [b,b] = [a,b] = [a,b^+] = 0.
\nom}
Using these commutation relations one can show that the Hamiltonian (\ref{1}) is in fact normal ordered w.r.t. $a, a^+, b, b^+$ operators. The operators $b, b^+$ correspond to ghosts. Let us introduce the parameter $\theta \equiv \frac{\,\,\omega_a}{\,\,\omega_b}$ and consider the eigenvalue problem
\disn{2}{
H |f\rangle = E |f\rangle.
\nom}
To find the behavior of the energy eigenvalues at small $\theta$ we decompose the Hamiltonian in powers of $\theta$ in the following way:
\disn{b42}{
H  =\omega_b (h_0 + \theta h_1), \quad \text{where} \quad h_0=-b^+b,\,\,\, h_1=a^+a + \frac{\lambda}{\,\,\omega_a}(a + a^+ + b + b^+)^4.
\nom}
The asymptotic expansions for the energy and the state vector are the following:
\disn{b43}{
E  = \omega_b (\e_0 + \theta \e_1 + \theta^2 \e_2 + \ldots), \quad \qquad |f\rangle = |f_0\rangle + \theta |f_1\rangle + \ldots
\nom}
To separate the energy eigenvalues that remain finite in the  $\theta \to 0$ limit we demand $\e_0=0$. Then in the lowest order approximation we get $h_0 |f_0\rangle = 0$, i.e. $b|f_0\rangle = 0.$ Therefore there are no ghost excitations in the state $|f_0\rangle$.
In the next order  we obtain the equation: $h_0 |f_1\rangle + h_1 |f_0\rangle = \e_1 |f_0\rangle$.
Taking the projection of this equation on the subspace $|f_0\rangle$ we obtain
the eigenvalue problem for the conventional and normal ordered anharmonic
oscillator Hamiltonian. The numerical solution of the problem (\ref{2})
confirms this result and makes it visual.
\begin{figure}
  \includegraphics[height=.3\textheight]{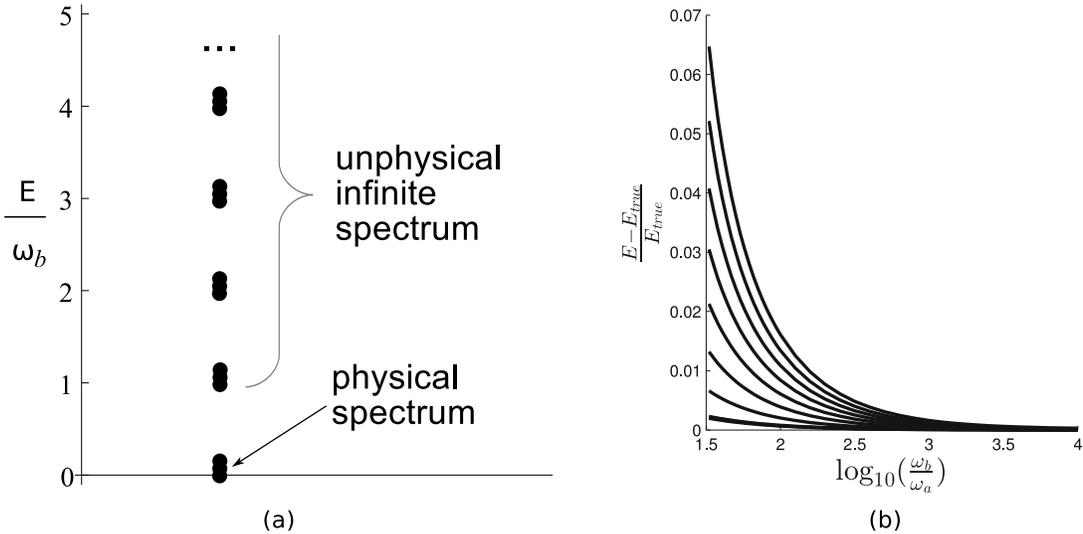}
  \caption{(\textbf{a}) The spectrum of the anharmonic oscillator with ghosts.
  (\textbf{b}) The convergence (at $\omega_b/\omega_a \to \infty$) of the first few eigenvalues of the spectrum,
  shown in (a), to the exact eigenvalues $E_{true}$ of the normal ordered anharmonic oscillator without ghosts}
 \label{oscillator}
\end{figure}

The spectrum of this theory looks like a series of equally-spaced sets of points (Fig.~\ref{oscillator}~(a)).
All but the first set approach infinity when  the parameter $\theta$ goes to zero (i.e. $\omega_b\to\infty$), so only
the first set corresponds to the spectrum of the anharmonic oscillator without ghosts. The convergence of the first few eigenvalues of this spectrum to the
exact eigenvalues of the normal ordered anharmonic oscillator without ghosts at $\theta \to 0$ is shown in Fig.~\ref{oscillator}~(b) (we denote these exact eigenvalues by $E_{true}$). The lowest curve in this figure corresponds to the first eigenvalue, the next curve -- to the second eigenvalue and so on.
One can see that lower eigenvalues approach faster to the limit.

\section{Conclusion}
\label{section:Conclusion}
In this paper we have considered the role of the PV regularization in field theory quantized on the LF. In section 2 we have reminded the idea of PV regularization using the example of scalar field theory with $\la\varphi^4$ interaction. In section 3 we have explained how PV regularization can help to remove possible inequivalence between perturbation theory generated by the quantization on the LF and usual covariant perturbation theory in Lorentz coordinates. We have considered the Yukawa-type model in (2+1) dimensions and showed how to construct the renormalized LF Hamiltonian perturbatively equivalent to the usually quantized theory using PV regularization.
In (3+1)-dimensional model we have no possibility to calculate exactly all counterterms necessary for the UV renormalization. So we have to consider these counterterms as terms with unknown couplings. This leads to difficulties in nonperturbative calculations with the LF Hamiltonian.

On the other hand, if one uses numerical methods in (2+1)-dimensional model, beside of the PV regularization, one has to reduce the number of field degrees of freedom. For example one can use the space lattice.  This lattice gives also UV regularization together with PV regularization. So we have the problem of removing these regularizations in nonperturbative calculations.

We have not considered examples of the application of PV regularization to gauge theories quantized on the LF. Such example was given for the QCD in \cite{NPPF, tmf99}. There the method for the constructing of the UV renormalized LF QCD Hamiltonian which generates the perturbation theory equivalent to the usual one in Lorentz coordinates was proposed. PV fields were used to restore this equivalence. Resulting LF QCD Hamiltonian contains many unusual counterterms necessary for UV renormalization. However it is difficult to use this Hamiltonian for nonperturbative (e.g. numerical) calculations due to the presence of many unknown couplings (i.e. coefficients before vertices related to these new counterterms).
So it could be useful to consider the example of (2+1)-dimensional QCD where these couplings can be calculated  owing to superrenormalizability of this model.
Also in gauge theories on the LF there is the problem of properly taking into account nonperturbative contributions (e.g. vacuum effects). This problem is rather complicated. As a variant of the solution of this problem in gauge theory on the LF one can apply the methods of DLCQ or semiphenomenological description of LF zero modes \cite{tmf2011}.

At the end of the present paper
we have given an example of calculations with PV-like ghosts and indefinite metric using the anharmonic oscillator which we modify by addition of ghost variables in analogy with PV regularization.
We have found that when the parameter analogous to large PV mass increases the part of the energy spectrum related to ghosts goes to infinity and switches off. The remaining part of the spectrum tends to the spectrum of ordinary anharmonic oscillator. We see that the lowest energy eigenvalues coincide very fast with its limit values.

\textbf{Acknowledgments.}
The authors thank the organizers of the International Conference "Quark Confinement and the Hadron Spectrum XI" 7-12 September 2014 in Saint Petersburg. Also the authors thank M.I. Vyazovsky for helpful discussions.
The authors
M.Yu. Malyshev, E.V. Prokhvatilov and R.A.~Zubov acknowledge Saint-Petersburg State University for a research grant
11.38.189.2014.


\end{document}